\documentclass[10pt,letterpaper]{article}

\usepackage{arxiv}

\usepackage[utf8]{inputenc} 
\usepackage[T1]{fontenc}    
\usepackage{hyperref}       
\usepackage{url}            
\usepackage{booktabs}       
\usepackage{amsfonts}       
\usepackage{nicefrac}       
\usepackage{microtype}      
\usepackage{lipsum}
\usepackage{graphicx}

\usepackage{comment}
\usepackage{multirow}
\usepackage{amssymb}
\usepackage{amsmath}
\usepackage{bbding}
\usepackage[table,xcdraw]{xcolor}

\author{
 Arth J. Shah \\
  Speech Research Lab\\
  Dhirubhai Ambani University (DAU)\\
  Gandhinagar, India \\
  \texttt{202521004@dau.ac.in} \\
   \And
 Devanshi K. Trivedi\\
  Sarvajanik College of Engineering \& Technology\\
  Surat, India \\
  \texttt{devanshitrivedi.co22d2@scet.ac.in} \\
  \And
 Himanshi U. Borad \\
  Sarvajanik College of Engineering \& Technology\\
  Surat, India \\
  \texttt{himanshiborad.co22d1@scet.ac.in} \\
   \And
 Hemant A. Patil \\
  Speech Research Lab \\
  Dhirubhai Ambani University (DAU) \\
  Gandhinagar, India \\
  \texttt{hemant\_patil@dau.ac.in} \\
}

\title{SingFox: A Multi-Lingual Singfake Detection Corpus}

\begin{document}
\maketitle

\begin{abstract}
    In this work, we introduce \textit{SingFox}, a comprehensive and large-scale dataset specifically to support robust evaluation of singing deepfake detection, and source tracing systems. SingFox is divided into six distinct tracks (T1–T6), each targeting a unique form of novelty—ranging from language diversity (global and Indian), genre-specific music, and alternative fake generations. The dataset encompasses over 113,802 audio clips across 20 languages, totaling more than 126.32 hours of audio data, and 1150 variety of singers. Each track is designed to imitate real-life situations and see how reliably the model performs under different conditions and to check its robustness. SingFox aims to foster reproducibility and accelerate research in singing deepfake detection by proposing a reliable dataset for both singfake detection task, and source verification task (explainability of model). We obtained highest accuracy of 77.84 \% while cross-testing models trained on various datasets. All the codes for reproducibility of dataset ara available at https://github.com/Arth-Shah/SingFox.

    \vspace{0.5cm}
    \textbf{Keywords:} SingFakes, Multi-Lingual, Alternative Fakes, Source Tracing

\end{abstract}

\begingroup
\renewcommand{\thefootnote}{}
\footnotetext{This manuscript is detailed version of SingFox paper, which has been accepted in INTERSPEECH 2026. Readers are requested to cite the published version of INTERSPEECH 2026.}
\addtocounter{footnote}{-1}

\section{Introduction}
In recent years, Artificial Intelligence (AI) and Machine Learning (ML) have enabled the generation of highly realistic synthetic content \cite{hong2025wildfake} \cite{yan2024df40} \cite{huang2025sida}. Since much progress has been made in the field of detecting synthetic images and videos, however, the field of Audio Deepfake Detection (ADD) remains shallowly explored. Within the audio domain, most of the researchers have concentrated on speech deepfakes, where AI generated/manipulated speech closely mimic a genuine speaker’s speech \cite{jung2025spoofceleb} \cite{li2025survey}. A recent and growing branch of this field is singing deepfakes (\textit{singfakes}) \cite{zang2024singfake}, where AI-generated models are trained to produce songs that mimics not only a singer’s timbre but also their characteristics, such as pitch (i.e., fundamental frequency, $F_0$), tone, and stylistic nuances. Unlike normal speech deepfakes, singfakes have additional musical and acoustic attributes, such as sustained rhythm, complex pitch ($F_0$) modulations, vibrato (slight variation in pitch during singing), and vocal melody. These attributes introduce significantly complex variability and structure, which makes singfake detection more challenging than normal speech-based deepfake detection \cite{zang2024singfake}. While singing synthesis and singing voice conversion technology, such as \cite{zhang2022visinger} and \cite{liu2022diffsinger} have advanced rapidly, they now also pose serious risks to security.  For example, an artist’s unique voice or musical style can be copied without consent, leading to copyright violations, misuse of reputation, or unauthorized commercial exploitation. Many a times, artist's voice is misused to promote products for sell or content they never agreed to. This raises significant \textit{ethical} and \textit{legal} concerns, particularly regarding personality rights, which protect an individual’s likeness, image, and voice.

Despite the growing threat of singfakes, the standard deepfake speech detection and speaker verification datasets, such as the ASVSpoof series \cite{wu2017asvspoof} \cite{todisco2019asvspoof} \cite{yamagishi2021asvspoof} \cite{WANG2026101825}, do not include singing deepfake audio. Alternatively, models trained via Deep Learning (DL) methods on speech-based datasets may fail to capture the singing attributes from speech, such as rhythm, melody, and harmonic characteristics in signals \cite{chen2024singing}. These consequences make singfake detection a comparatively shallow and emerging research problem, several studies have attempted to adapt ASVSpoof datasets for singing deepfake detection \cite{zang2024singfake}. However, as reported in \cite{zhang2024svdd}, these approaches yield limited performance due to the mismatch between speech-centric training data and singing-specific artifacts based testing data. Table \ref{cmp_datasets} summarizes currently available datasets for singing deepfake detection. Although several resources exist, each exhibits notable limitations. For example, SONICS, the largest dataset to date with 4,751 hours of data \cite{RahmanHSPF25}, relies exclusively on two generation architectures, i.e., Udio and Suno AI \cite{casini2025data}. This limited architectural diversity restricts its ability to represent singfakes generated using alternative paradigms, such as Generative Adversarial Networks (GANs) and/or Voice Conversion (VC) techniques. Language diversity characteristics, i.e., multilingualism also  pose another key limitation while testing. Most existing datasets are based on single language, i.e., English, which questions models implementation to real-life scenarios, where input can be a speech in any language. While CtrSVDD \cite{CtrSVDD} includes two languages (Chinese and Japanese) and WildSVDD contains six languages \cite{zang2024singfake}, multilingual representation remains insufficient for evaluating real-world robustness. 
\begin{table*}[h!]
\centering
\scriptsize
\tiny
\caption{Comparison of Different Existing Datasets \textit{w.r.t.} Proposed Dataset (Utt = utterances, Alt = Alternative)}
\label{cmp_datasets}
\scalebox{0.78}
{\begin{tabular}{|
>{\columncolor[HTML]{FFFFFF}}c |
>{\columncolor[HTML]{FFFFFF}}c |
>{\columncolor[HTML]{FFFFFF}}c |
>{\columncolor[HTML]{FFFFFF}}c |
>{\columncolor[HTML]{FFFFFF}}c |
>{\columncolor[HTML]{FFFFFF}}c |
>{\columncolor[HTML]{FFFFFF}}c |
>{\columncolor[HTML]{FFFFFF}}c |
>{\columncolor[HTML]{FFFFFF}}c |
>{\columncolor[HTML]{FFFFFF}}c |
>{\columncolor[HTML]{FFFFFF}}c |
>{\columncolor[HTML]{FFFFFF}}c |
>{\columncolor[HTML]{FFFFFF}}c |
>{\columncolor[HTML]{FFFFFF}}c |}
{\color[HTML]{24292E} \textbf{Dataset}}                      & {\color[HTML]{24292E} \textbf{\begin{tabular}[c]{@{}c@{}}\# \\ Lang.\end{tabular}}} & \textbf{\begin{tabular}[c]{@{}c@{}}\# \\ Music\end{tabular}} & \textbf{\begin{tabular}[c]{@{}c@{}}\# \\ Utt\end{tabular}} & \textbf{\# Real} & \textbf{\# Fake} & \textbf{\begin{tabular}[c]{@{}c@{}}\# \\ Alt. \\ Fake\end{tabular}} & \textbf{GAN} & \textbf{\begin{tabular}[c]{@{}c@{}}DM\end{tabular}} & \textbf{VC} & \textbf{TTM} & \textbf{\begin{tabular}[c]{@{}c@{}}\#\\Singer\end{tabular}} & \textbf{\begin{tabular}[c]{@{}c@{}}\# Total \\ Hour\end{tabular}} & \textbf{\begin{tabular}[c]{@{}c@{}}Text \\ Symm\\ etric\end{tabular}} \\ \hline
FSD \cite{xie2024fsd}                                                         & 1                                                                                       & 650                                                           & 650                                                               & 200              & 450                                                                               & X                                                                           & X            & 1                                                                    & 4           & X            & 12+                 & 26.26                                                              & X                                                                     \\ \hline
WildSVDD  \cite{zhang2024svdd}                                                   & 6                                                                                       & X                                                             & 28,205                                                            & 12,751           & 15,454                                                                             & 28,205                                                                      & X            & 1                                                                    & 1           & X            & 97                  & 58.33                                                              & X                                                                     \\ \hline
CtrSVDD \cite{CtrSVDD}                                                     & 2                                                                                       & X                                                             & 220,798                                                           & 32,312           & 188,486                                                                           & X                                                                           & X            & X                                                                    & 7           & X            & 164                 & 307.98                                                             & X                                                                     \\ \hline
FMC \cite{comanducci2025fakemusiccaps}                                                         & 1                                                                                       & 27,605                                                        & 27,605                                                            & 5,500            & 22,105                                                                            & X                                                                           & X            & 4                                                                    & X           & 1            & -                   & 77                                                                 & \Checkmark                                                                     \\ \hline
SONICS \cite{RahmanHSPF25}                                                      & 1                                                                                       & X                                                             & 97,164                                                            & 48,090           & 49,074                                                                            & X                                                                           & X            & 2                                                                    & X           & 2            & 9k+               & 4,751                                                              & \Checkmark                                                                     \\ \hline
\begin{tabular}[c]{@{}c@{}}SingFox\\ (Proposed)\end{tabular} & \textbf{20}                                                                             & 12,725                                                       & 113,802                                                           &  54,911          & 37,953                                                                    & 10,469                                                                      & 3            & 2                                                                    & 2           & 1            & 1150  & 126.32                                                           & \Checkmark                                                                     \\ \hline
\end{tabular}}

\end{table*}

The aim of training a model is to capture the characteristics of generative models; however, when evaluating the model's robustness, we need a real-world, scenario-specific dataset. Although monolingual datasets may be sufficient during model training, as the primary objective is to capture AI-generated artifacts, real-world attacks are neither restricted to a single language nor confined to known generation methods, therefore testing models within these limitations and questioning their generalization and real-life deployment case. Real-life attacks that involve diverse languages and unseen synthesis techniques make the task even more challenging. For speech deepfake, many dataset, such as In-The-Wild \cite{muller2022does}, Multi-Language Audio Anti-spoofing Dataset (MLAAD) \cite{muller2024mlaad}, and ASVSpoof 2021 \cite{ASVSpoof_2021LA} have notable made progress and have been widely adopted for testing, however, there still exist no such datasets in singing deepfake field. To alter this limitation, this study propose a novel dataset, namely, SingFox, which contains 20 languages (14 international, and 6 Indic). Furthermore, existing singfake datasets lack diversity of generation models diversity, as most of them rely on a single class of models, such as diffusion models (DMs), Voice Conversion (VC), Text-to-Music (TTM) systems, or GAN-based vocoders, without integrating multiple paradigms within a single dataset. Such diversification is critical for improving robustness and explainability, which also further helps in identifying artifacts accurately. In simple words, a detection model trained (bias) on DM-based generated singfakes may under-perform when evaluated against VC-generated singfakes, and vice-versa. Testing on proposed SingFox helps to investigate, this paradigm, and also to provide \textit{explainability} of models behavior. Employing multiple diverse synthesis techniques within a dataset enables stronger cross-model generalization, and more accurate results (models behavior in real-life scenarios).

Another critical concern while training and testing is \textit{text asymmetry}. Several existing datasets exhibit misalignment between the transcribed lyrics and the generated audio, where models can be biased to detect non-vocabulary words. Text symmetry is essential for generalized model development and testing, as it encourages the model to learn meaningful differences in vocoder artifacts and acoustic modeling characteristics between genuine and spoofed samples, rather than relying on irrelevant factors, such as gender variation, linguistic imbalance, or phoneme distribution. Singfox includes 1150 singers, and enabling robust evaluation across diverse vocal characteristics, which exhibits in real-life scenarios. Since this study do not focus on speaker verification problem, and due to copyright issues, the label of speaker identity has not been disclosed. By employing multilingual dataset, multiple synthesis generation models, and text-symmetric design, the dataset establishes a more realistic and challenging benchmark for singfake detection $^{\pm}$ \footnote{Dataset available for non-commertial research purpose only at $^{\pm}$\textit{https://doi.org/10.5281/zenodo.20691932}.}.

This study offers the following key contributions:
\begin{itemize}
    \item Multi-lingual In-The-Wild SingFakes for testing models generalization.
    \item Source tracing, and explainability for singfakes.
    \item Alternative Singfakes (fake vocals and real background music ).
    \item GANS, diffusion models, TTMs, and VC-based singfake generation (Anonymous demo available at $\mp$\footnote{\textit{$\mp$ https://shorturl.at/Sa1M7}}).
\end{itemize}


\section{Dataset Generation}
\label{2}
\subsection{Real Audio}
\label{2.1}
Recordings of audios from singers is time consuming and costly. Therefore, we downloaded songs in various languages from different open source websites, which provide copyright free songs to be downloaded, such as at $https://pixabay.com/music/$. In order to eliminate possible \textit{bias} in \textit{found data} to build proposed corpus, we selected a few keywords and searched relevant audios from the website. The audios were then converted to standard "\textit{.flac}" format, and were resampled at \textit{16 kHz} sampling rate in order to maintain uniformity across the entire dataset. To avoid shortcut biases (e.g., the model relying on uneven volume levels or silent gaps in fake songs instead of detecting genuine deepfake traces), we applied dual normalization on each audio sample: (i) peak normalization, which adjusts the maximum amplitude of the signal to a target-level, and (ii) Root Mean Square (RMS) normalization, which adjusts the average loudness (signal energy) to a fixed RMS-level. The processed audios thereafter were randomly queued and trimmed into a smaller chunks of 4 seconds each for all 20 languages. The 4 seconds protocol was motivated via standard SingFake dataset \cite{zang2024singfake}. Furthermore, to avoid additional sources of \textit{bias}, care was taken to ensure that there is no overlap in singers, languages, or data throughout the dataset. Since the dataset is aimed to resemble real-life scenario case, and not only training scenario, no source separation is applied to separate source music from vocals. Fig. \ref{fig:preprocessing} displays the steps followed for preprocessing of audios before generation of singfakes from audios. 
\begin{figure}[h!]
    \centering
    \includegraphics[width=\linewidth]{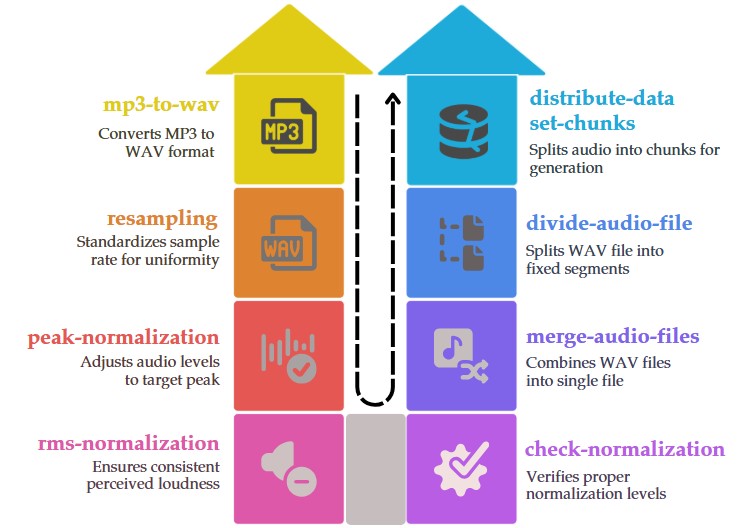}
    \caption{Steps applied for pre-processing of the dataset.}
    \label{fig:preprocessing}
\end{figure}
Table \ref{tab:stat} displays the statistical difference between total number of speakers per language. The gender of the speaker, and genre distribution remains anonymous. 
\begin{table}[h!]
\scriptsize
\centering
\caption{Speaker statistics of real audio \{ * = Language codes based on standard ISO code \cite{byrum1999iso}\} }
\label{tab:stat}
\begin{tabular}{|
>{\columncolor[HTML]{FFFFFF}}c |
>{\columncolor[HTML]{FFFFFF}}c |
>{\columncolor[HTML]{FFFFFF}}c |
>{\columncolor[HTML]{FFFFFF}}c |
>{\columncolor[HTML]{FFFFFF}}c |
>{\columncolor[HTML]{FFFFFF}}c |}
\hline
\textbf{no.} & Language* & \# Singers & no. & Language* & \# Singers \\ \hline
1            & en       & 67       & 11  & pt       & 62       \\ \hline
2            & hi       & 65       & 12  & id       & 65       \\ \hline
3            & fr       & 60       & 13  & ja       & 70       \\ \hline
4            & de       & 58       & 14  & tr       & 55       \\ \hline
5            & es       & 55       & 15  & vi       & 43       \\ \hline
6            & it       & 63       & 16  & bn       & 50       \\ \hline
7            & ko       & 70       & 17  & ta       & 45       \\ \hline
8            & zh       & 69       & 18  & pa       & 48       \\ \hline
9            & ar       & 64       & 19  & te       & 45       \\ \hline
10           & ru       & 51       & 20  & mr       & 45       \\ \hline
\end{tabular}
\end{table} 
\begin{figure*} [h!]
    \centering
    \includegraphics[width=\linewidth]{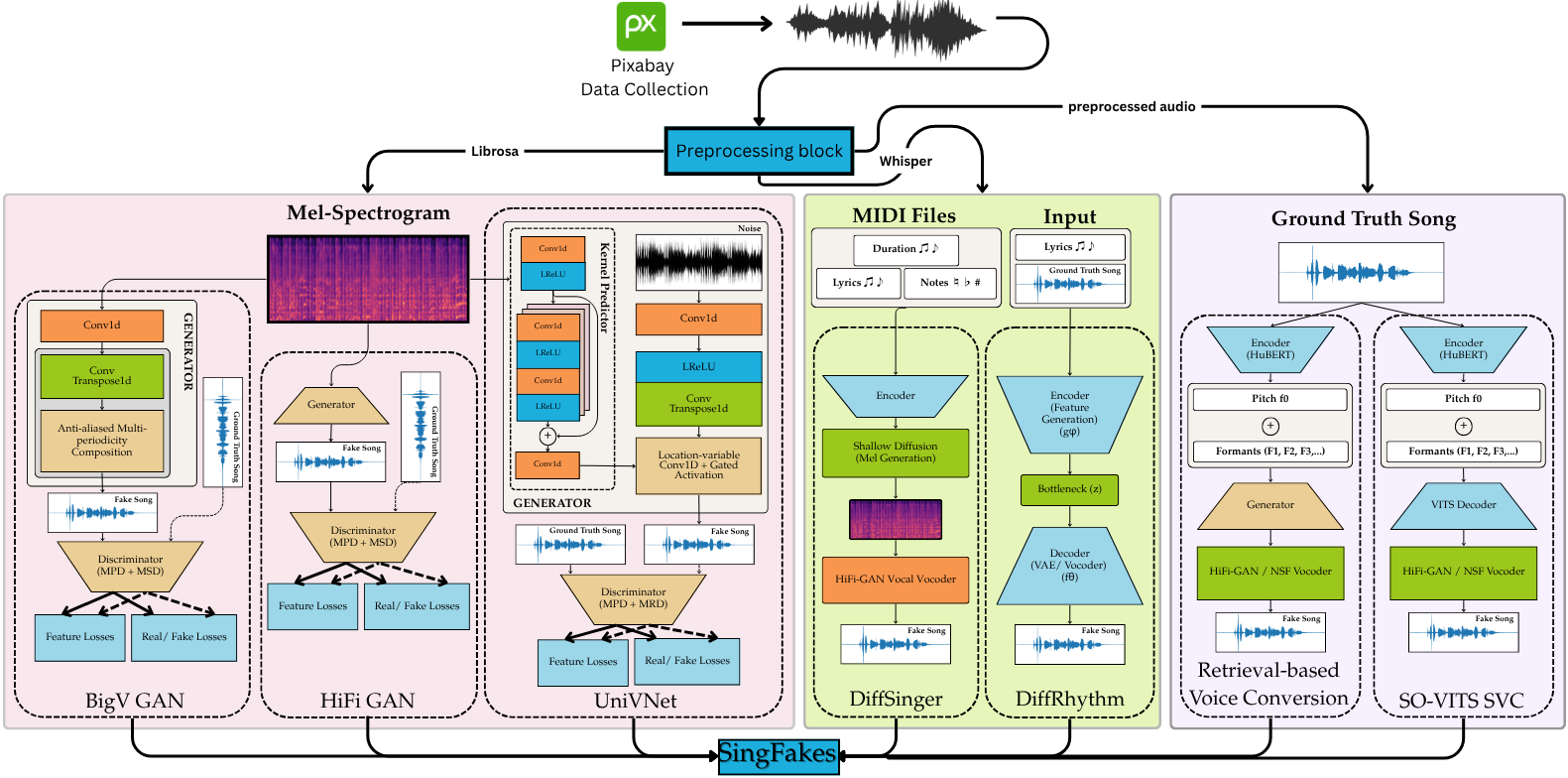}
    \caption{End-to-end SingFox pipeline illustrating data collection, preprocessing, singfake generation, and ground-truth modeling using multiple vocoders and diffusion-based frameworks.}
    \label{processing}
\end{figure*}
\subsection{Singfake Audios}
We generated singfakes via 4 different methods, namely, GANs, DMs, VC models, and TTM models. Particularly, we employed 3 different types of GANs, 2 different types of DMs, 2 different VCs, and 1 TTM, which are supported by numerous open source models to generate fake singing utterances. Not all the models took raw waveforms as input (except VCs), where some models take lyrics, duration, and notes of songs as input, for which we used OpenAI's pretrained Whisper-960h large model to generate the text transcripts \cite{radford2023robust} (with average WER of 4.9 \% (of selected languages)). GANs took Mel spectrograms as input to generate fake audios, which were generated using librosa library of python. We downloaded AI-generated audios, which are generated using TTM-based (MusicGen \cite{han2024understanding}) model. 

\subsubsection{Models Used to Generate Singfakes}
We employ 3 GAN-based vocoders—HiFi-GAN (Universal) \cite{kong2020hifi}, BigVGAN \cite{lee2023bigvgan}, and UnivNet \cite{JangLYKK21}—to synthesize fake singing voice from Mel spectrograms. HiFi-GAN provides real-time, high-quality generation. BigVGAN enhances robustness and fidelity with larger capacity and anti-aliased activations; and UnivNet achieves competitive performance with lightweight design. Together, they enable diverse and comprehensive evaluation of GAN vocoders. We use 2 diffusion models, DiffSinger \cite{liu2022diffsinger}, and DiffRhythm \cite{ning2025diffrhythm}, for symbolic-to-audio generation. DiffSinger converts MIDI-based \textit{phonetic }and \textit{musical} inputs into expressive Mel spectrograms, while DiffRhythm models rhythm and timing for musically coherent synthesis. Both use HiFi-GAN vocoders for waveform generation, offering controllable, high-quality, and expressive singfake audio. For voice conversion, we adopt Retrieval-based VC (RVC) \cite{RVC} and So-VITS-SVC. Both use HuBERT embeddings \cite{hsu2021hubert} and pitch ($F_0$) features to preserve \textit{melody} and \textit{rhythm} while converting timbre. RVC combines linguistic features with pitch/formant information, while So-VITS-SVC decodes them using a VITS-based architecture. These VC models generate natural, identity-preserving singfakes with a minimal distortion. All the models were pretrained on universal data, which consists of several renounced dataset, such as VCTK \cite{yamagishi2019cstr}, LibriTTS \cite{panayotov2015librispeech}, and LJ Speech \cite{ljspeech17}, thereby able to generate songs on multiple languages. Detailed codes and block diagram for reproducibility are available on $^{\bigoplus}$ \footnote{$^{\bigoplus}$https://github.com/Arth-Shah/SingFox}. Fig. \ref{processing} displays an detailed diagram of pipeline to generate singfakes (end-to-end colab notebooks are available at github).

\subsection{Novelty in Each Track}
We designed the dataset with a real-life testing scenarios in mind, therefore, we decided to design a separate track depending on several user-specific conditions. Track T1 covers the 14 most spoken languages in the world (except Indic languages), thereby making a globally generalized dataset sub-track. Since India is a linguistically diverse nation, it has over 1,536 languages and dialects ($\approx$ one-fourth of the world’s languages), and Indic languages are severely underrepresented in audio deepfake research. We were motivated to generate a separate track for the Indic language, which resulted in a rare multilingual, non-Western evaluation benchmark. Moreover, the risk associated with each language in deepfake detection is equally significant \cite{mai2023warning}. Several existing datasets (as reported in Table \ref{cmp_datasets}) have proposed a variety of deepfakes (non-vocal instrumental), however, they are limited to a few musical instruments. Hence, to alleviate this limitation, we propose 5 different types of instrument-based deepfakes in Track T3. Track T4 is a super-track of T1 and T2, resulting in the answer to the question \textit{"Would the model be valid for singfake detection globally?"}. Authors strongly encourage researchers to test the generated model on Track T4, unless they aim to develop a model for a specific region (e.g., Indic or European languages). Visual demonstration of languages distribution is represented in Fig. \ref{fig:language}.
\begin{figure}
    \centering
    \includegraphics[width=\linewidth]{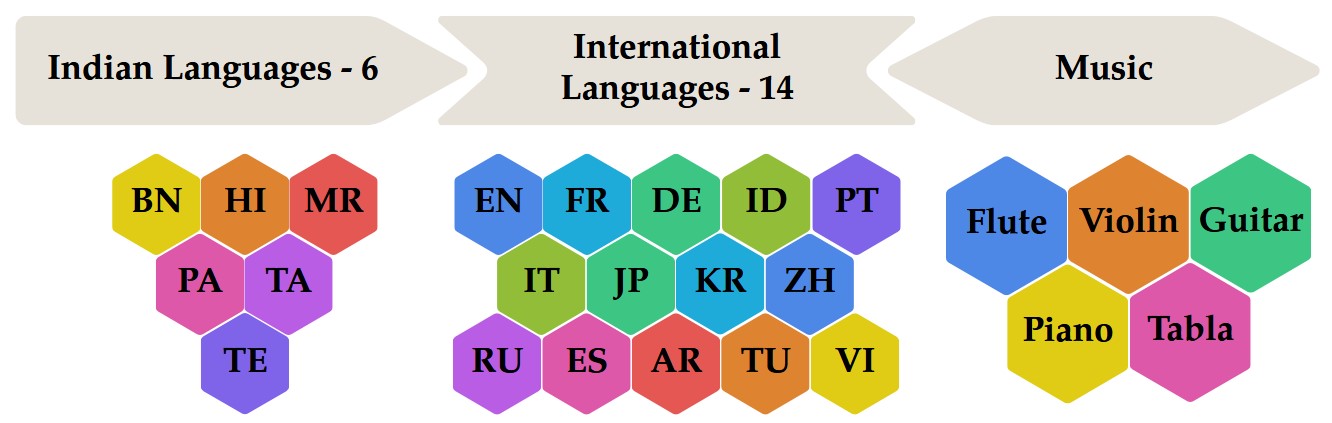}
    \caption{Musics Languages in SingFox Dataset as per ISO code.}
    \label{fig:language}
\end{figure}

\subsubsection{Track T5: Alternative Fakes}
Alternative fakes are basically fake vocals of a singer in a song, and real background music, which confuses the system and helps attackers to conduct an attack successfully. As per our best knowledge and belief, \textit{SingFox} is the first corpus that proposes alternative fakes for 20 different languages. Random instrumental real audios were mixed (vocals and background music were simply mixed together to form a single audio signal. The mixed signal was then converted into mono format so that both components exist in one single-channel audio) with a few vocals, and were analyzed in three different classes, i.e, real vocals + real instruments (12,011 trials), fake vocals + fake instruments (10,469 trials), and real instruments + fake vocals (10,469 trials). Since the attacks, where the vocals are real (singers identity and authenticity are preserved), regardless of whether instruments are real or fake, that particular scenario was not accounted for. Fig. \ref{fig:T5_models} displays the number of files used from each model in each dataset while generating alternative fake audios.

\begin{figure}[h!]
    \centering
    \includegraphics[width=\linewidth]{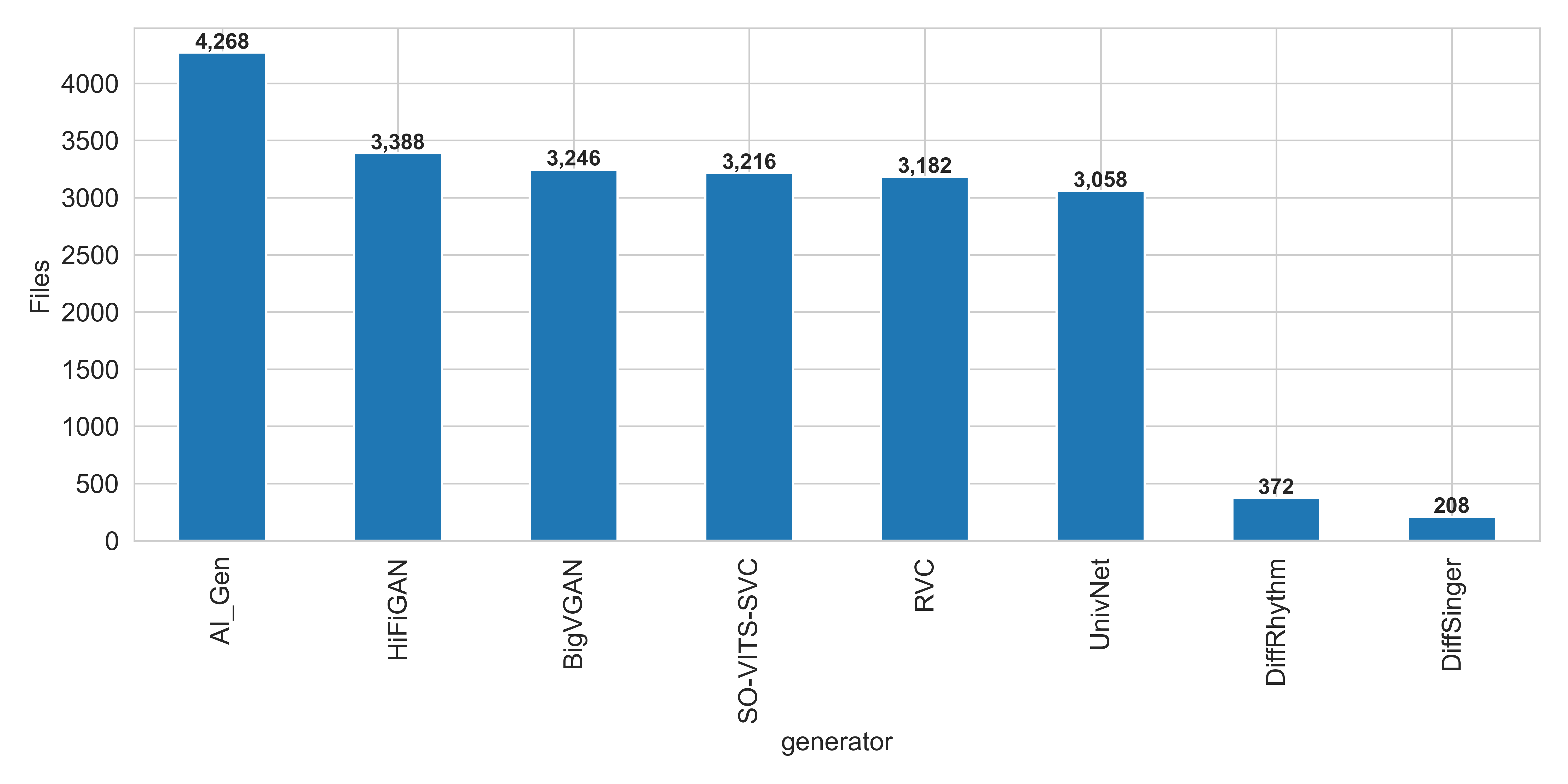}
    \caption{Model-wise distribution of generated singfakes in T5 track.}
    \label{fig:T5_models}
\end{figure}

\subsubsection{Track T6: Source Tracing}
Since \textit{Source Tracing} (ST) is a new field in the area of deepfake detection, which basically supports the \textit{explainability} of models performance and helps to identify what properties model fails to capture, there exist no works that supports the explainability of generated singing deepfakes \cite{st1} \cite{st2} \cite{st3}. Motivated by this, authors propose a new track to detect from which source has the audio been generated. For this particular study, we apply Source Verification approach to track the source of generated audio \cite{sv1, sv2, sv3}. No model overlap condition for enrollment and evaluation was applied (with similar condition as reported in \cite{sv2}). Since ST system is never used independently, and always employed to enhance the singfake detection system, we can employ pre-trained deepfake detection system model for source verification, as reported in \cite{sv2}. For this particular study, we employ LFCC + ResNet trained on SingFox (proposed) dataset for setup and baseline system, however, authors highly motivate researchers to use pretrained model on WildSVDD \cite{zang2024singfake} dataset, which provides standard training data. Since we do not need a separate set of dataset for ST, only a protocol file (containing three columns) on testing set of Track T4, which is used in experiments. The proposed source tracing framework consists of two stages: \textit{enrollment} and \textit{evaluation}.
\begin{itemize}
\item{\textbf{Enrollment Stage: }}
For each synthetic speech generation model (source), a set of enrollment utterances is collected to construct the source profile. Let $S_i$ denote a source model. The enrollment set for each source contains 20 speech samples:

\begin{equation}
S_i = \{\texttt{file}_1, \texttt{file}_2, \ldots, \texttt{file}_{20}\}.
\end{equation}

For example, the enrollment protocol includes:

\begin{itemize}
    \item HiFiGAN: $\{\texttt{file}_1, \ldots, \texttt{file}_{20}\}$
    \item SO-VITS-SVC: $\{\texttt{file}_1, \ldots, \texttt{file}_{20}\}$
    \item DiffRhythm: $\{\texttt{file}_1, \ldots, \texttt{file}_{20}\}$
\end{itemize}

\item{\textbf{Evaluation Stage: }}
During evaluation, each test utterance is paired with a claimed source identity. The evaluation tuple is represented as: (claim\_source, file\_name, label), where the label is either \texttt{positive\_source} or \texttt{negative\_source}.

For a test sample generated by HiFiGAN, the evaluation protocol is:

\begin{verbatim}
HiFiGAN      abc_1.flac    positive_source
SO-VITS-SVC  abc_2.flac    negative_source
DiffRhythm   abc_3.flac    negative_source
\end{verbatim}

Here, the claimed source matches the true generating model only for HiFiGAN, resulting in a positive trial, while all other source claims form negative trials.

\item{\textbf{Unseen Source Evaluation: }}
To evaluate open-set robustness, utterances generated by unseen synthesis models are included. Since these sources are absent from the enrollment stage, all source claims are treated as negative trials. For example:

\begin{verbatim}
HiFiGAN      abc_1.flac     negative_source
SO-VITS-SVC  abc_32.flac     negative_source
DiffRhythm   abc_1.flac     positive_source
\end{verbatim}

where \texttt{BigVGAN} is an unseen source not present in the enrollment database.

\item{\textbf{Evaluation Scenarios: }}
The resulting evaluation protocol contains two categories of trials:

\begin{itemize}
    \item \textbf{Seen-source trials}: one positive trial corresponding to the true source and multiple negative trials corresponding to other enrolled sources.
    \item \textbf{Unseen-source trials}: only negative trials, since the true source is not enrolled.
\end{itemize}

This protocol enables the assessment of:

\begin{enumerate}
    \item \textbf{Closed-set source tracing}, which measures the ability to correctly attribute speech to one of the enrolled sources.
    \item \textbf{Open-set source rejection}, which evaluates the capability of rejecting utterances generated by previously unseen sources.
    \item \textbf{Forensic attribution robustness}, which examines the reliability of source attribution under realistic deployment conditions involving both known and unknown synthesis models.
\end{enumerate}
\end{itemize}
In the case when user wish to perform source tracing protocols for the task different then from source verification \cite{sv1}, users are free to generate their own protocols from file name as they are well annotated. Language and music of all the files in SingFox dataset can be found in Fig. \ref{fig:language_distribution}. 
\begin{figure}[h!]
    \centering
    \includegraphics[width=\linewidth]{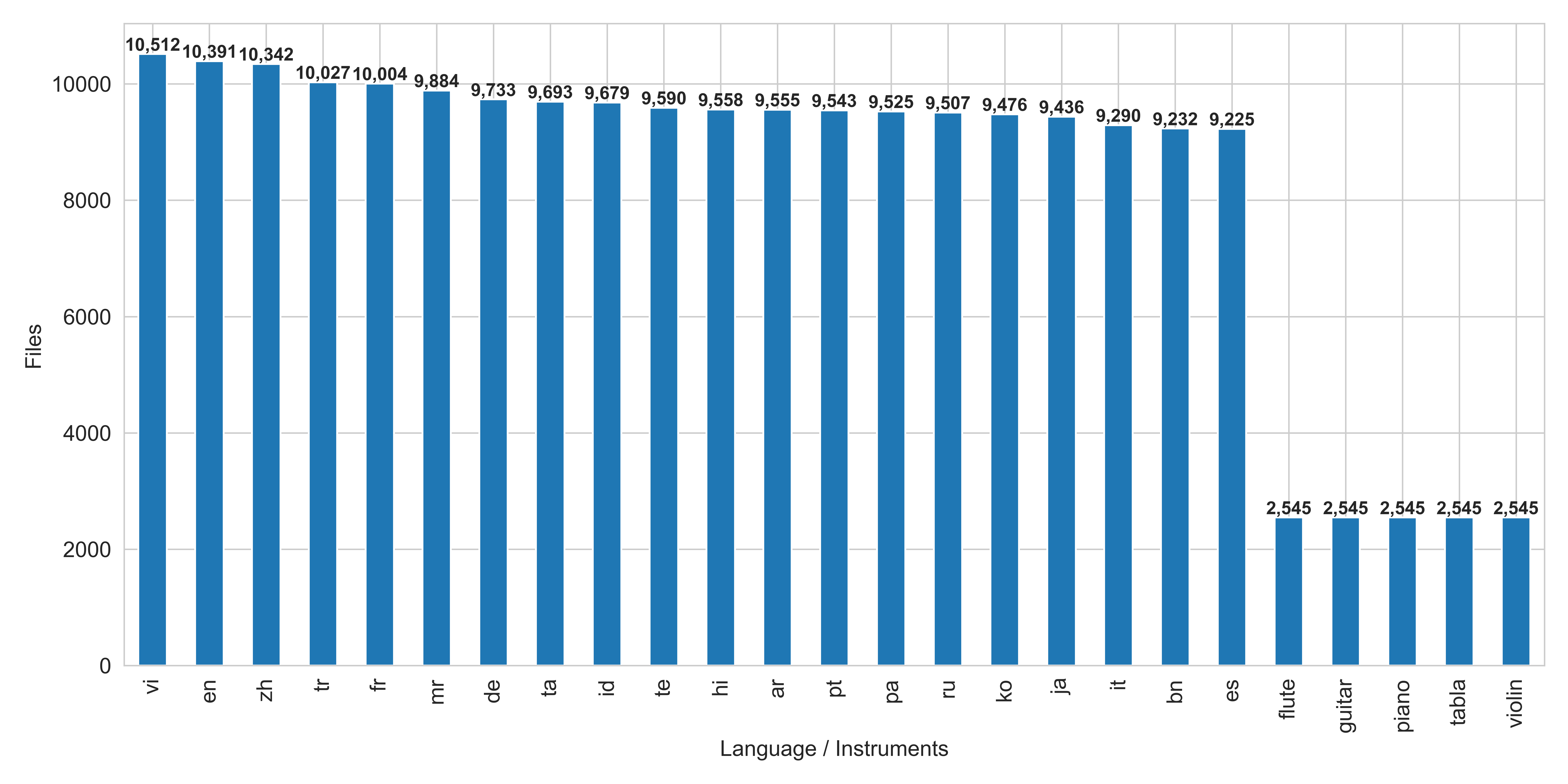}
    \caption{Language and music distribution in SingFox dataset.}
    \label{fig:language_distribution}
\end{figure}

The distribution of models across each datasets, i.e., which models were used for generating singfakes in which track can be found in Fig. \ref{fig:model_distribution}.
\begin{figure}[h!]
    \centering
    \includegraphics[width=\linewidth]{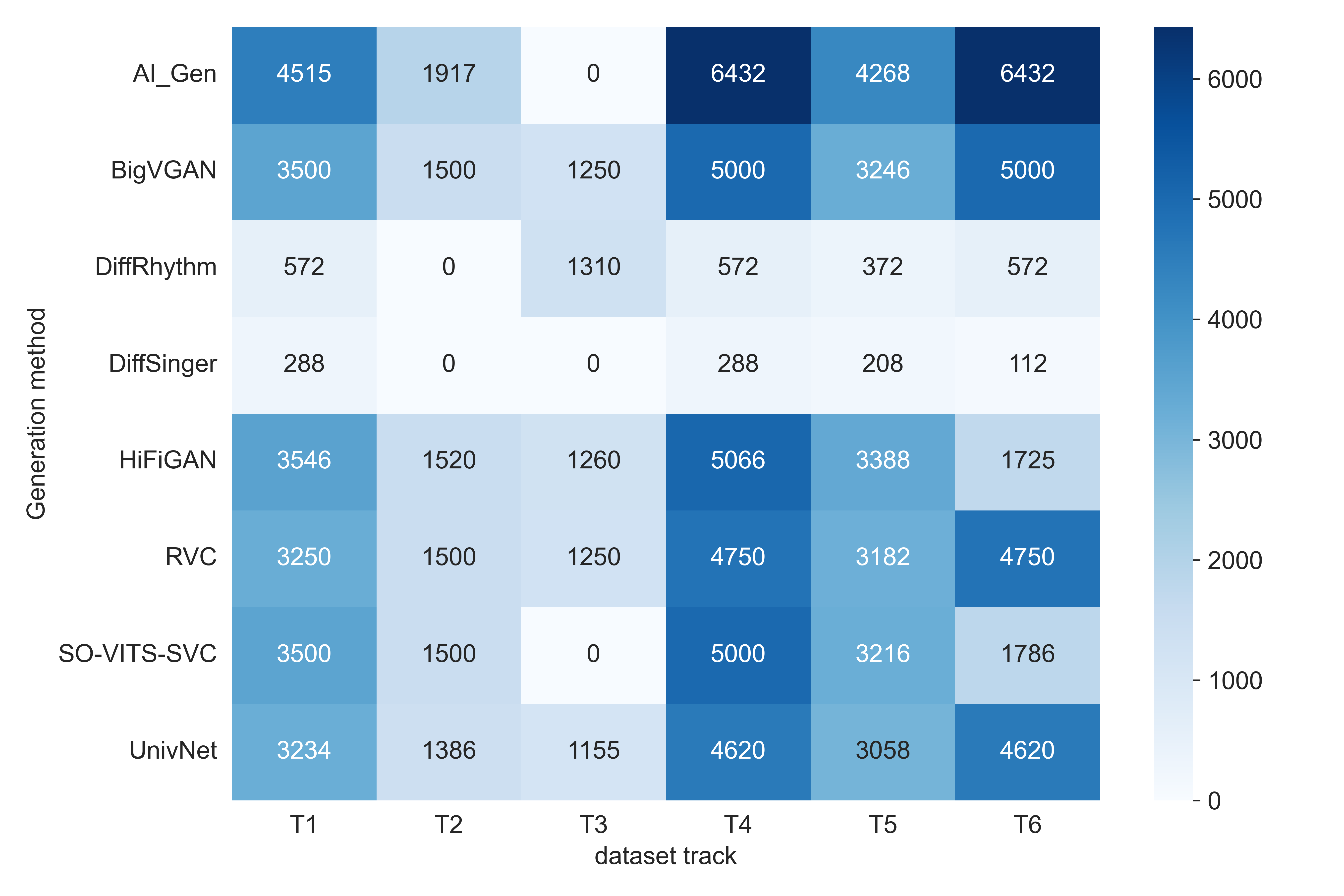}
    \caption{Model distribution per dataset track.}
    \label{fig:model_distribution}
\end{figure}

A small subset ($\approx$ 30 \%) of each track of dataset was used for training, where only HiFi-GAN, SO-VITS-SVC, and DiffRhythm were used while training, and validation, while remaining model (including a sub-portion of models used for training) were used for testing. Train / test split can be observed in Fig. \ref{tt_split}. 
\begin{figure}
    \centering
    \includegraphics[width=\linewidth]{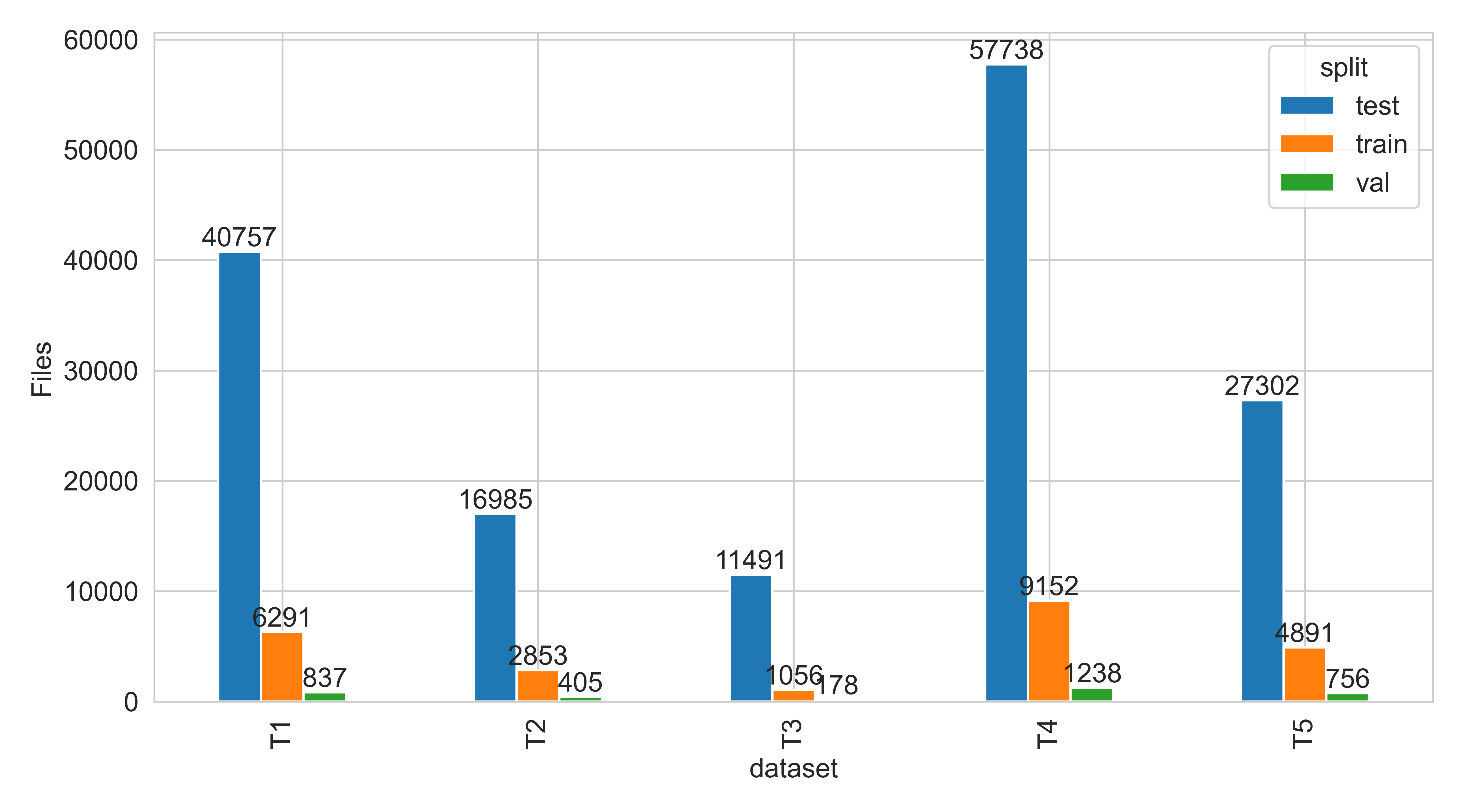}
    \caption{Train/Val/testing distribution across different tracks of dataset.}
    \label{tt_split}
\end{figure} 

Table \ref{tracts} presents the statistics of data for different tracks of SingFox dataset.

The detailed distribution of number of files per language can observed in Appendix \ref{appendix_a}. 

\begin{table}[h!]
\scriptsize
\centering
\caption{Comparison (statistics) of Different Tracks of SingFox Dataset \{* both, fake vocals + real music, and fake vocals + fake music, $^{+}$ denotes music types\}}
\label{tracts}
\scriptsize
\begin{tabular}{|c|c|c|c|c|c|c|}
\hline
                                                           Track$=>$            & T1     & T2     & T3     & T4     & T5     & T6     \\ \hline
\begin{tabular}[c]{@{}c@{}}Total\\ Files\end{tabular}                  & 47,885 & 20,243 & 12,725 & 68,128 & 32,949 & 24,997 \\ \hline
Lang.                                                                  & 14     & 6      & 0 ($5^{+}$)      & 20     & 20     & 20     \\ \hline
Singfakes                                                              & 22,405 & 9,323  & 6,225  & 31,728 & 20,938* & 24,997 \\ \hline
\begin{tabular}[c]{@{}c@{}}Size \\ (in GB)\end{tabular}                & 4.3    & 1.89   & 0.89   & 6.18   & 3.05   & 2.21   \\ \hline
\begin{tabular}[c]{@{}c@{}}Total\\ duration\\ (in hrs.)\end{tabular}   & 53.17  & 22.47  & 14.12  & 75.63  & 36.57  & 27.73  \\ \hline
\begin{tabular}[c]{@{}c@{}}Avg.\\ duration\\ per lang.\end{tabular} & 227.86 & 224.67 & 169.40 & 226.90 & 109.70 & 83.20  \\ \hline
\end{tabular}
\end{table}
\section{Experiments}
For experimental validation of proposed dataset, we explored various acoustic features, as well as Self-Supervised Learning (SSL)-based features to provide a baseline system. We primarily selected spectral features, such as Linear Frequency Cepstral Coefficients (LFCC) + ResNet as a baseline system, which was also one of four baseline system explored in \cite{zhang2024svdd}. In addition, experiments are performed with various other acoustic features, such as MFCC \cite{davis1980comparison}, and GFCC \cite{zhao2013analyzing}, in combination with various classifiers, such as CNN \cite{lecun2015deep}, BiLSTM \cite{siami2019performance}, and BiGRU \cite{cho2014learning}. 

Fig. \ref{spider_web_chart} shows the \textit{radar plot} of obtained results for acoustic features on T1 to T5 track, from which it can be observed that when the number of languages increases, the robustness of models increases and also accuracy increases. In case of T5, it can be observed that the lowest accuracy observed is 45.13 \% on LFCC+ResNet classifier, where this sharp drop in accuracy denotes the difficulty in model distinguishing between fake vocals and real music. Table \ref{tab:ST} represents the robustness of LFCC feature (with ResNet classifier) in detecting spoofed audios properties, whereas GFCC fails to identify type of spoof file (i.e., GFCC model has \textit{weak explainability} as compared to LFCC model). More detailed information about evaluation metrices is available in Appendix \ref{appendix_b}
\begin{figure}[h!]
    \centering    
    \includegraphics[width=0.6\linewidth]{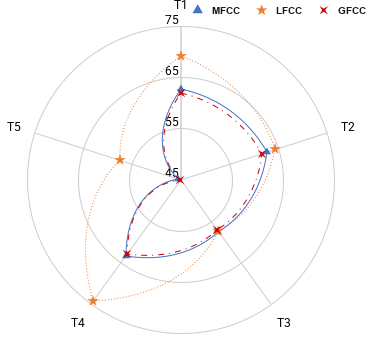}
    \caption{Results on various acoustic features with ResNet classifier on various tracks of proposed singfox dataset.}
    \label{spider_web_chart}
\end{figure}
\begin{table}[h!]
\centering
\caption{Source tracing performance on ResNet classifier}
\label{tab:ST}
\begin{tabular}{|
>{\columncolor[HTML]{FFFFFF}}l |
>{\columncolor[HTML]{FFFFFF}}l |
>{\columncolor[HTML]{FFFFFF}}l |
>{\columncolor[HTML]{FFFFFF}}l |}
\hline
Features =\textgreater{} & MFCC                                               & LFCC                                                        & GFCC                                               \\ \hline
Accuracy (in \%)         & \multicolumn{1}{r|}{\cellcolor[HTML]{FFFFFF}88.71} & \multicolumn{1}{r|}{\cellcolor[HTML]{FFFFFF}\textbf{89.06}} & \multicolumn{1}{r|}{\cellcolor[HTML]{FFFFFF}70.34} \\ \hline
\end{tabular}
\end{table}
To validate dataset on SOTA methods, we performed SSL-based experiments on T4 track (largest track), whose Detection Error Trade-off (DET) curve can be observed from Fig. \ref{DET_curve}. It can be seen that acoustic features, such as LFCC (with BiLSTM as backend classifier) outperforms SSL-based Wav2Vec2 features and almost all the SOTA models, excluding Raw2Net. This phenomena likely happens due to small training data of proposed SingFox dataset, which can be altered by training models on WildSVDD dataset \cite{zang2024singfake}.
\begin{figure}[h!]
    \centering
    \includegraphics[width=\linewidth]{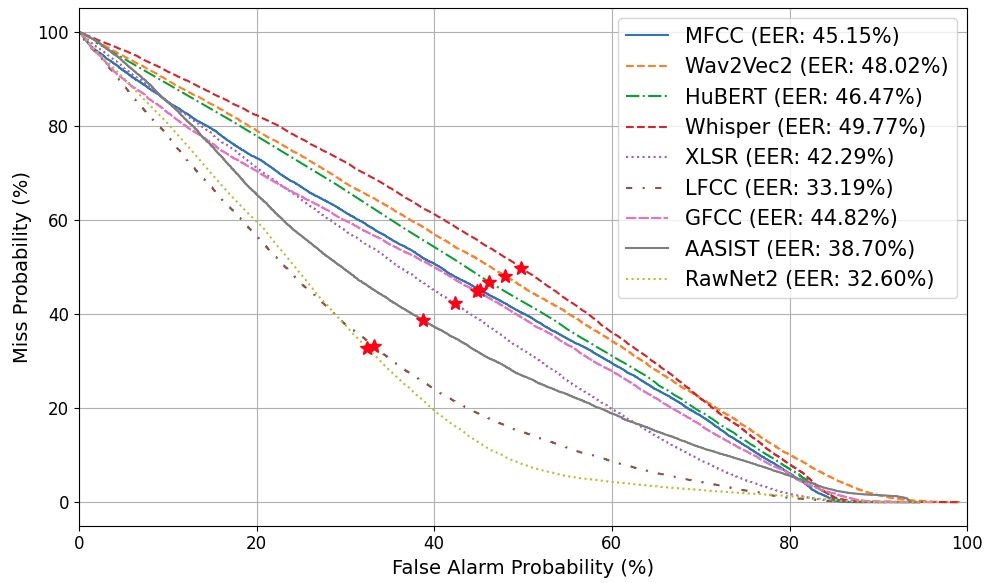}
    \caption{DET curve (\# singfake trials = 24,997, \# genuine trials = 32,741) of T6 \textit{w.r.t.} different systems.}
    \label{DET_curve}
\end{figure}
\subsection{Cross-Dataset Results}
In order to validate the testing generalization of the SingFox dataset, we performed \textit{cross-dataset} experiments. Table \ref{cross_dataset} shows the results on three of the existing singfake datasets. Other datasets, such as FSD and SONICS, where either not open source, or were too large to be accommodated due to resource constraints, which restricted us to perform experiments at much larger scale. For conducting cross-dataset experiments, LFCC + ResNet was selected as baseline system for all the tracks, with similar configurations as in \cite{zhang2024svdd}. It can be observed that when the model is trained on CtrSVDD dataset, or WildSVDD dataset, it performs poorly on SingFox dataset as compared to FMC dataset. This phenomena can primarily occur due to robustness of data and variety of data available for training in FMC dataset. Table \ref{cross_dataset} demonstrate how much good a model would perform in real-life scenario. Since we do not propose SingFox for training, and have limited training data, we in this section illustrate how the models trained on different datasets perform in real-life scenarios, i.e., by validating them against SingFox testing data. It can be observed that when the model is trained on CtrSVDD dataset, or WildSVDD dataset, the model misperforms due to generalizability issues.  However, when the model is trained on FMC dataset, the model gives robust performance across multi-lingual dataset denoting its robust behaviour and generalizibility. 

\begin{table}[h!]
\small
\caption{Results (Acc. in \%) Under Cross-Dataset Scenarios}
\centering
\label{cross_dataset}
\resizebox{\columnwidth}{!}{
\begin{tabular}{|
>{\columncolor[HTML]{FFFFFF}}c |
>{\columncolor[HTML]{FFFFFF}}c |
>{\columncolor[HTML]{FFFFFF}}c |
>{\columncolor[HTML]{FFFFFF}}c |}
\hline
\cellcolor[HTML]{FFFFFF}{\color[HTML]{000000} \textbf{\begin{tabular}[c]{@{}c@{}}Train → \\\hline  Test ↓\end{tabular}}} & {\color[HTML]{24292E} CtrSVDD} & WildSVDD & FMC         \\ \hline
CtrSVDD                                                                                   & 65.87                          & 43.88    & 36.13\\ \hline
WildSVDD                                                                                  & 44.65                          & \textbf{71.55}    & 49.11\\  \hline
FMC                                                                                       & \textbf{77.58}                 & 22.24    & \textbf{98.32} \\ \hline
SingFox (T4)                                                                                  & 46.06                          &  54.17   & 77.84 \\ \hline
\end{tabular}}
\end{table}
\subsection{Model Specific and Human-Centric Evaluation}
In order to evaluate the quality of generated singfakes via different deepfake models, we conducted experiments based on how much accurate a model works on specific singfake generation technique? Table \ref{tab:human_evaluation} displays the results of model-wise accuracy, i.e., how much accurate does the baseline system (the best in Fig. \ref{cmp_datasets}) works on a particular model? Since the models used for training gave almost accurate performance (due to low account of training data), they were replaced with "X" in Table \ref{tab:human_evaluation}. It can be observed that UniVNet works the best giving an accuracy of 71.17 \%, and BigVGAN works worst (signifies highly accurate singfake generation) giving an accuracy of 1.02 \%. 
\begin{table*}[h!]
\centering
\scriptsize
\caption{Objective Evaluation and Model-Specific Accuracy (in \%) (performance metrics described in Appendix \ref{appendix_b}).}
\label{tab:human_evaluation}
\resizebox{\columnwidth}{!}{
\begin{tabular}{|c|c|c|c|c|c|c|c|}
\hline
\textbf{Models} & \textbf{\begin{tabular}[c]{@{}c@{}}PESQ\\ \cite{pesq} (↑)\end{tabular}} & \textbf{\begin{tabular}[c]{@{}c@{}}STOI\\ \cite{stoi}(↑)\end{tabular}} & \textbf{\begin{tabular}[c]{@{}c@{}}PCC\\  \cite{pcc}(↑)\end{tabular}} & \textbf{\begin{tabular}[c]{@{}c@{}}MSD\\  \cite{msd}(↓)\end{tabular}} & \textbf{\begin{tabular}[c]{@{}c@{}}MCD\\  \cite{MCD}(↓)\end{tabular}} & \multicolumn{1}{c|}{\textbf{\begin{tabular}[c]{@{}c@{}}MOS\\ \cite{mos} (↑)\end{tabular}}} & \textbf{\begin{tabular}[c]{@{}c@{}}Model\\ Acc. (↑)\end{tabular}} \\ \hline
\textbf{HiFiGAN} & 1.08 & 1.9E-1 & 8.6E-2 & 4.1E-1 & 498.73 & 2.00 & X \\ \hline
\textbf{BigVGAN} & 3.18 & 9.7E-1 & 9.4E-1 & \textbf{5.0E-5} & \textbf{23.06} & 4.12 & 1.02 \\ \hline
\textbf{DiffSinger} & 1.04 & 1.1E-2 & 6.7E-2 & 3.2E-4 & 427.38 & 3.80 & X \\ \hline
\textbf{DiffRhythm} & 1.04 & -1.9E-2 & 1.9E-1 & 5.2E-4 & 241.31 & \textbf{4.39} & 58.39 \\ \hline
\textbf{\begin{tabular}[c]{@{}c@{}}SO-VITS-\\ SVC\end{tabular}} & 1.05 & 3.5E-1 & 1.0E-1 & 8.4E-4 & 251.23 & 3.51 & X \\ \hline
\textbf{RVC} & \textbf{4.64} & \textbf{1.0E+0} & \textbf{1.0E+0} & 1.4E-3 & 44.43 & 3.84 & 1.77 \\ \hline
\textbf{UniVNet} & 1.19 & 1.8E-1 & 5.6E-1 & 2.7E-4 & 86.30 & 2.61 & \textbf{71.17} \\ \hline
\end{tabular}} 
\end{table*}

We conducted subjective human-centric experiments in order to evaluate quality of audios perceptual (51 non-native English speakers were selected, aged between 19 and 25 years, with no known hearing impairments). For known languages only participants were asked to rate authenticity on a \textit{5-}point Mean Opinion Score (MOS) scale. Since we could not accumulate listeners of all 20 different languages, perceptual validation for quality of some languages still remains open research problem (so does language specific MOS). The generated deepfakes (averaged for speaker known language) achieved a MOS of 3.468 (indicates the good quality of generated deepfakes), closely approaching the MOS 4.028 of ground-truth audios. Both objective and subjective assessment metrics are used to thoroughly evaluate the quality of generated speech across several models, as shown in Table 4. PESQ, STOI, PCC, MSD, MCD, MOS, and model-wise classification accuracy are among the measures. In audio processing, objective evaluation is a conventional methodology. Particularly well-known standards for evaluating speech quality and intelligibility include PESQ and STOI, which have been frequently used in significant speech enhancement tasks as the CHiME-7 UDASE task.

The high perceptual quality of a signal is generally indicated by a PESQ score that is closer to 4.5. With the highest PESQ of 4.64 among the models, RVC demonstrated exceptional audio fidelity. BigVGAN, which also performed well, and achieved a PESQ of 3.18. STOI is used to show the differences in intelligibility between two signals. RVC and BigVGAN demonstrated exceptional speech intelligibility with STOI values (~1.0 and 0.97, respectively). In terms of correlation (PCC), RVC leads with a high correlation of 1.0, while BigVGAN closely follows (0.94). This implies that the outputs of these models are quite similar to the reference signals.
The spectral similarity is revealed by the MSD values, which showed that BigVGAN had the lowest MSD (5.0E-5), followed by UniVNet and DiffSinger. These findings show that the generated and reference modulation spectra for these models are very similar. A lower number for MCD means that it is more closely aligned with ground truth. The lowest MCD values, BigVGAN (23.06) and RVC (44.43) demonstrated their capacity to preserve high-quality spectral characteristics.

It is crucial to remember that, when it comes to model-wise classification accuracy, a lower accuracy means that the generated deepfakes is more realistic because the classifier finds it difficult to discriminate between generated deepfakes and ground-truths. Despite RVC having higher objective scores, the subjective MOS evaluations of participants showed that DiffRhythm (4.39) and BigVGAN (4.12) were the most favored. This draws attention to a significant trade-off between human perception and objective measurements. These results give us observations that features that are acoustically significant may not be perceptually relevant. 

\section{Summary and Conclusions}
This study developed \textit{SingFox} corpus-a novel multilingual singing voice dataset comprising six tracks (T1-T6), each designed to facilitate the development and evaluation of deepfake detection in \textit{singing} audio. With over 113,802 files spanning across 20 different languages, 5 different music types, and 126.32 hours of audio, SingFox introduces diverse, high-resolution content for evaluation of SOTA models. Furthermore, we aim to perform source separation-based experiments along with separate multi-lingual training datasets for each of the proposed tracks. Exploring variety of cross-attention-based classifiers and newly proposed methods, such as SingGraph \cite{chen2024singing} remains an open research problem for singfake detection task . 

\section{Acknowledgement}

The authors sincerely thank Mr. Neil Zhang and the Audio Information Research Lab, University of Rochester, USA for providing access to the WildSVDD dataset. We also acknowledge the creators of the SingFake, WildSVDD, and Ctrl-SVDD corpora for making these valuable resources available to the research community, which greatly supported this work.

\section{Generative AI Disclosure}

Generative AI tools were used for language polishing and bug fixes during the preparation of this work. All generated outputs were carefully reviewed, verified, and approved by the authors, who take full responsibility for the final content.

\bibliography{library}

@inproceedings{radford2023robust,
  title={Robust speech recognition via large-scale weak supervision},
  author={Radford, Alec and Kim, Jong Wook and Xu, Tao and Brockman, Greg and McLeavey, Christine and Sutskever, Ilya},
  booktitle={International Conference on Machine Learning (ICML)},
  pages={28492--28518},
  year={2023, Honolulu, HI, USA},
  organization={}
}

@inproceedings{han2024understanding,
  title={{Understanding the use of AI-based audio generation models by end-users}},
  author={Han, Jiyeon and Yang, Eunseo and Oh, Uran},
  booktitle={Extended Abstracts of the CHI Conference on Human Factors in Computing Systems},
  volume = {355},
  pages        = {1--7},
  publisher    = {{ACM}},
  year={2024, Hamburg, Germany}
}

@inproceedings{xie2024fsd,
  title={{FSD: An initial {Chinese} dataset for fake song detection}},
  author={Xie, Yuankun and Zhou, Jingjing and Lu, Xiaolin and Jiang, Zhenghao and Yang, Yuxin and Cheng, Haonan and Ye, Long},
  booktitle={IEEE International Conference on Acoustics, Speech and Signal Processing (ICASSP)},
  pages={4605--4609},
  year={2024, Seoul, Korea},
  organization={}
}

@inproceedings{zhang2024svdd,
  title={{SVDD 2024: The inaugural singing voice deepfake detection challenge}},
  author={Zhang, You and Zang, Yongyi and Shi, Jiatong and Yamamoto, Ryuichi and Toda, Tomoki and Duan, Zhiyao},
  booktitle={IEEE Spoken Language Technology Workshop (SLT)},
  pages={782--787},
  year={2024, Macao, China},
  organization={}
}

@inproceedings{CtrSVDD,
  author       = {Yongyi Zang and
                  Jiatong Shi and
                  You Zhang and
                  Ryuichi Yamamoto and
                  Jionghao Han and
                  Yuxun Tang and
                  Shengyuan Xu and
                  Wenxiao Zhao and
                  Jing Guo and
                  Tomoki Toda and
                  Zhiyao Duan},
  title        = {{CtrSVDD: {A} benchmark dataset and baseline analysis for controlled
                  singing voice deepfake detection}},
  booktitle    = {INTERSPEECH 2024, Kos, Greece},
pages={4783-4787},
year={}
}

@article{comanducci2025fakemusiccaps,
  title={Fakemusiccaps: A dataset for detection and attribution of synthetic music generated via text-to-music models},
  author={Comanducci, Luca and Bestagini, Paolo and Tubaro, Stefano},
  journal={Journal of Imaging},
  volume={11},
  number={7},
  pages={242},
  year={2025},
  publisher={MDPI}
}

@inproceedings{RahmanHSPF25,
  author       = {Md. Awsafur Rahman and
                  Zaber Ibn Abdul Hakim and
                  Najibul Haque Sarker and
                  Bishmoy Paul and
                  Shaikh Anowarul Fattah},
  title        = {{SONICS:} {S}ynthetic Or Not - Identifying Counterfeit Songs},
  booktitle    = {The $13^{th}$ International Conference on Learning Representations,
                  {(ICLR)}, Singapore},
  year         = {2025}
}

@inproceedings{yamagishi2021asvspoof,
  title={{ASVSpoof 2021: Accelerating progress in spoofed and deepfake speech detection}},
  author={Yamagishi, Junichi and Wang, Xin and Todisco, Massimiliano and Sahidullah, Md and Patino, Jose and Nautsch, Andreas and Liu, Xuechen and Lee, Kong Aik and Kinnunen, Tomi and Evans, Nicholas and others},
  booktitle={ASVSpoof Workshop},
  pages={47-54},
  year={2021, Kos Island, Greece}
}

@article{kong2020hifi,
  title={{HiFi-GAN: Generative adversarial networks for efficient and high-fidelity speech synthesis}},
  author={Kong, Jungil and Kim, Jaehyeon and Bae, Jaekyoung},
  journal={Advances in Neural Information Processing Systems (NIPS), Virtual},
  volume={33},
  pages={17022--17033},
  year={2020}
}

@inproceedings{lee2023bigvgan,
  title={{BigVGAN: {A} universal neural vocoder with large-scale training}},
  author={Lee, Sang Gil and Ping, Wei and Ginsburg, Boris and Catanzaro, Bryan and Yoon, Sungroh},
  booktitle={$11^{th}$ International Conference on Learning Representations (ICLR), Kigali, Rwanda},
  year={2023}
}

@inproceedings{JangLYKK21,
  author       = {Won Jang and
                  Dan Lim and
                  Jaesam Yoon and
                  Bongwan Kim and
                  Juntae Kim},
  title        = {{UnivNet: {A} neural vocoder with multi-resolution spectrogram discriminators
                  for high-fidelity waveform generation}},
  booktitle    = {INTERSPEECH},
  pages        = {2207--2211},
  year         = {2021, Brno, Czechia}
}

@article{ning2025diffrhythm,
  title={DiffRhythm: Blazingly fast and embarrassingly simple end-to-end full-length song generation with latent diffusion},
  author={Ning, Ziqian and Chen, Huakang and Jiang, Yuepeng and Hao, Chunbo and Ma, Guobin and Wang, Shuai and Yao, Jixun and Xie, Lei},
  journal={arXiv preprint arXiv:2503.01183},
  year={2025 \{Last Accessed: $17^{th}$ Feb, 2026\}}}

@article{RVC,
    author = {RVC-Project},
    title = {{GitHub: https://github.com/RVC-Project/Retrieval-based-Voice-Conversion-WebUI}},
    journal = {\{Last Accessed: $27^{th}, February, 2025$\}},
    year = {2024}
}

@article{hsu2021hubert,
  title={{HuBERT: Self-supervised speech representation learning by masked prediction of hidden units}},
  author={Hsu, Wei-Ning and Bolte, Benjamin and Tsai, Yao-Hung Hubert and Lakhotia, Kushal and Salakhutdinov, Ruslan and Mohamed, Abdelrahman},
  journal={IEEE/ACM Transactions on Audio, Speech, and Language Processing},
  volume={29},
  pages={3451--3460},
  year={2021},
  publisher={}
}

@article{davis1980comparison,
  title={Comparison of parametric representations for monosyllabic word recognition in continuously spoken sentences},
  author={Davis, Steven and Mermelstein, Paul},
  journal={IEEE Transactions on Acoustics, Speech, and Signal Processing},
  volume={28},
  number={4},
  pages={357--366},
  year={1980},
  publisher={}
}

@inproceedings{zhao2013analyzing,
  title={{Analyzing noise robustness of MFCC and GFCC features in speaker identification}},
  author={Zhao, Xiaojia and Wang, DeLiang},
  booktitle={IEEE International Conference on Acoustics, Speech, and Signal Processing (ICASSP)},
  pages={7204--7208},
  year={2013, Vancover, Canada},
  organization={}
}

@article{mai2023warning,
  title={Warning: Humans cannot reliably detect speech deepfakes},
  author={Mai, Kimberly T and Bray, Sergi and Davies, Toby and Griffin, Lewis D},
  journal={PLoS One},
  volume={18},
  number={8},
  pages={285--333},
  year={2023},
  publisher={Public Library of Science}
}

@inproceedings{huang2025sida,
  title={{SIDA}: Social media image deepfake detection, localization, and explanation with large multimodal model},
  author={Huang, Zhenglin and Hu, Jinwei and Li, Xiangtai and He, Yiwei and Zhao, Xingyu and Peng, Bei and Wu, Baoyuan and Huang, Xiaowei and Cheng, Guangliang},
  booktitle={Computer Vision and Pattern Recognition Conference (CVPR)},
  pages={28831--28841},
  year={2025, Nashville, Tennessee, USA}
}

@article{yan2024df40,
  title={Df40: Toward next-generation deepfake detection},
  author={Yan, Zhiyuan and Yao, Taiping and Chen, Shen and Zhao, Yandan and Fu, Xinghe and Zhu, Junwei and Luo, Donghao and Wang, Chengjie and Ding, Shouhong and Wu, Yunsheng and others},
  journal={Advances in Neural Information Processing Systems (NeurIPS)},
  volume={37},
  pages={29387--29434},
  year={2024, Vancouver, Canada}
}

@inproceedings{hong2025wildfake,
  title={Wildfake: A large-scale and hierarchical dataset for {AI}-generated images detection},
  author={Hong, Yan and Feng, Jianming and Chen, Haoxing and Lan, Jun and Zhu, Huijia and Wang, Weiqiang and Zhang, Jianfu},
  booktitle={Proceedings of the AAAI Conference on Artificial Intelligence},
  volume={39},
  number={4},
  pages={3500--3508},
  year={2025, Philadelphia, Pennsylvania, USA}
}

@article{li2025survey,
  title={A survey on speech deepfake detection},
  author={Li, Menglu and Ahmadiadli, Yasaman and Zhang, Xiao-Ping},
  journal={ACM Computing Surveys},
  volume={57},
  number={7},
  pages={1--38},
  year={2025},
  publisher={ACM New York, NY}
}

@article{jung2025spoofceleb,
  title={{SpoofCeleb}: Speech deepfake detection and {SASV} in the wild},
  author={Jung, Jee-weon and Wu, Yihan and Wang, Xin and Kim, Ji-Hoon and Maiti, Soumi and Matsunaga, Yuta and Shim, Hye-jin and Tian, Jinchuan and Evans, Nicholas and Chung, Joon Son and others},
  journal={IEEE Open Journal of Signal Processing},
  year={2025},
  volume={6},
  pages={68--77},
  publisher={IEEE}
}

@inproceedings{zang2024singfake,
  title={Singfake: Singing voice deepfake detection},
  author={Zang, Yongyi and Zhang, You and Heydari, Mojtaba and Duan, Zhiyao},
  booktitle={IEEE International Conference on Acoustics, Speech and Signal Processing (ICASSP)},
  pages={12156--12160},
  year={2024, Seoul, Korea},
  organization={}
}

@inproceedings{zhang2022visinger,
  title={Visinger: Variational inference with adversarial learning for end-to-end singing voice synthesis},
  author={Zhang, Yongmao and Cong, Jian and Xue, Heyang and Xie, Lei and Zhu, Pengcheng and Bi, Mengxiao},
  booktitle={IEEE International Conference on Acoustics, Speech and Signal Processing (ICASSP)},
  pages={7237--7241},
  year={2022, (Virtual) Singapore},
  organization={}
}

@inproceedings{liu2022diffsinger,
  title={Diffsinger: Singing voice synthesis via shallow diffusion mechanism},
  author={Liu, Jinglin and Li, Chengxi and Ren, Yi and Chen, Feiyang and Zhao, Zhou},
  booktitle={AAAI conference on Artificial Intelligence},
  volume={36},
  number={10},
  pages={11020--11028},
  year={2022, (Virtual) USA}
}

@article{wu2017asvspoof,
  title={{ASVSpoof}: The automatic speaker verification spoofing and countermeasures challenge},
  author={Wu, Zhizheng and Yamagishi, Junichi and Kinnunen, Tomi and Hanil{\c{c}}i, Cemal and Sahidullah, Mohammed and Sizov, Aleksandr and Evans, Nicholas and Todisco, Massimiliano and Delgado, Hector},
  journal={IEEE Journal of Selected Topics in Signal Processing},
  volume={11},
  number={4},
  pages={588--604},
  year={2017},
  publisher={}
}

@inproceedings{todisco2019asvspoof,
  title={{ASVSpoof} 2019: Future Horizons in Spoofed and Fake Audio Detection},
  author={Todisco, Massimiliano and Wang, Xin and Vestman, Ville and Sahidullah, Md and Delgado, H{\'e}ctor and Nautsch, Andreas and Yamagishi, Junichi and Evans, Nicholas and Kinnunen, Tomi and Lee, Kong Aik},
  booktitle={INTERSPEECH},
  pages={1008--1012},
  year={2019, Graz, Austria}
}

@article{WANG2026101825,
title = {{ASVspoof} 5: Design, collection and validation of resources for spoofing, deepfake, and adversarial attack detection using crowdsourced speech},
journal = {Computer Speech \& Language},
volume = {95},
pages = {101825},
year = {2026},
issn = {0885-2308},
doi = {https://doi.org/10.1016/j.csl.2025.101825},
author = {Xin Wang and Héctor Delgado and Hemlata Tak and Jee-weon Jung and Hye-jin Shim and Massimiliano Todisco and Ivan Kukanov and Xuechen Liu and Md Sahidullah and Tomi Kinnunen and Nicholas Evans and Kong Aik Lee and Junichi Yamagishi and Myeonghun Jeong and Ge Zhu and Yongyi Zang and You Zhang and Soumi Maiti and Florian Lux and Nicolas Müller and Wangyou Zhang and Chengzhe Sun and Shuwei Hou and Siwei Lyu and Sébastien {Le Maguer} and Cheng Gong and Hanjie Guo and Liping Chen and Vishwanath Singh}
}

@inproceedings{chen2024singing,
  title={Singing Voice Graph Modeling for SingFake Detection},
  author={Chen, Xuanjun and Wu, Haibin and Jang, Roger and Lee, Hung-yi},
  booktitle={INTERSPEECH},
  pages={4843--4847},
  year={2024, Kos Island, Greece}
}

@article{casini2025data,
  title={Data-Driven Analysis of Text-Conditioned {AI}-Generated Music: A Case Study with Suno and Udio},
  author={Casini, Luca and Vila, Laura Cros and Dalmazzo, David and Kaila, Anna-Kaisa and Sturm, Bob LT},
  journal={arXiv preprint arXiv:2509.11824},
  year={2025, \{Last Accessed: $27^{th} February, 2026$\}}
}

@article{ASVSpoof_2021LA,
  title={{ASVSpoof} 2021: Towards spoofed and deepfake speech detection in the wild},
  author={Liu, Xuechen and Wang, Xin and Sahidullah, Md and Patino, Jose and Delgado, H{\'e}ctor and Kinnunen, Tomi and Todisco, Massimiliano and Yamagishi, Junichi and Evans, Nicholas and Nautsch, Andreas and others},
  journal={IEEE/ACM Transactions on Audio, Speech, and Language Processing},
  volume={31},
  pages={2507--2522},
  year={2023},
  publisher={}
}

@inproceedings{muller2022does,
  title={Does Audio Deepfake Detection Generalize?},
  author={M{\"u}ller, Nicolas and Czempin, Pavel and Diekmann, Franziska and Froghyar, Adam and B{\"o}ttinger, Konstantin},
  booktitle={INTERSPEECH},
  pages={2783--2787},
  year={2022, Incheon, Korea}
}

@inproceedings{muller2024mlaad,
  title={{MLAAD: The multi-language audio anti-spoofing dataset}},
  author={M{\"u}ller et al., Nicolas M},
  booktitle={International Joint Conference on Neural Networks (IJCNN), Yokohama, Japan},
  pages={1--7},
  year={2024},
  organization={}
}

@inproceedings{byrum1999iso,
  title={ISO 639-1 and ISO 639-2: International Standards for Language Codes. ISO 15924: International Standard for Names of Scripts.},
  author={Byrum, John D},
  booktitle = {IFLA Council and General Conference},
  year={1999, Bangkok, Thailand},
  publisher={ERIC}
}

@article{yamagishi2019cstr,
  title={{CSTR VCTK Corpus}: {English} multi-speaker corpus for CSTR voice cloning toolkit (version 0.92)},
  author={Yamagishi, Junichi and Veaux, Christophe and MacDonald, Kirsten},
  journal={The Rainbow Passage which the speakers read out can be found in the International Dialects of English Archive:(http://web. ku. edu/\~{} idea/readings/rainbow. htm).},
  year={2019},
  publisher={University of Edinburgh. The Centre for Speech Technology Research (CSTR)}
}

@inproceedings{panayotov2015librispeech,
  title={Librispeech: an asr corpus based on public domain audio books},
  author={Panayotov, Vassil and Chen, Guoguo and Povey, Daniel and Khudanpur, Sanjeev},
  booktitle={IEEE International Conference on Acoustics, Speech and Signal Processing (ICASSP)},
  pages={5206--5210},
  year={2015, South Brisbane, Queensland, Australia},
  organization={}
}

@misc{ljspeech17,
  author       = {Keith Ito and Linda Johnson},
  title        = {The {LJ Speech} Dataset},
  howpublished = {\url{https://keithito.com/LJ-Speech-Dataset/}},
  year         = 2017,
  note = {\{Last Accessed: $3^{rd}$ March, 2026\}}
}

@inproceedings{st1,
  title={Source Tracing of Audio Deepfake Systems},
  author={Klein, Nicholas and Chen, Tianxiang and Tak, Hemlata and Casal, Ricardo and Khoury, Elie},
  booktitle={INTERSPEECH},
  pages={1100--1104},
  year={2024, Kos Island, Greece}
}

@inproceedings{st2,
  title={Source tracing: Detecting voice spoofing},
  author={Zhu, Tinglong and Wang, Xingming and Qin, Xiaoyi and Li, Ming},
  booktitle={Asia-Pacific Signal and Information Processing Association Annual Summit and Conference (APSIPA ASC)},
  pages={216--220},
  year={2022, Chiang Mai, Thailand},
  organization={}
}

@inproceedings{st3,
  title={Multilingual Source Tracing of Speech Deepfakes: A First Benchmark},
  author={Xuan, Xi and Xiao, Yang and Das, Rohan Kumar and Kinnunen, Tomi},
  booktitle={$5^{th}$ Symposium on Security and Privacy in Speech Communicatio (SPSC)},
  pages={27--34},
  year={2025, Delft, Netherlands}
}

@incollection{sv1,
  title={Source verification for Speech Deepfakes},
  author={Negroni, V and Salvi, D and Bestagini, P and Tubaro, S and others},
  booktitle={INTERSPEECH},
  pages={1--5},
  year={2025, Rotterdam, Netherlands}
}

@inproceedings{sv2,
  title={{STOPA}: A Dataset of Systematic VariaTion Of DeePfake Audio for Open-Set Source Tracing and Attribution},
  author={Firc, Anton and Chhibber, Manasi and Mishra, Jagabandhu and Pratap Singh, Vishwanath and Kinnunen, Tomi and Malinka, Kamil},
  booktitle={INTERSPEECH},
  pages={1553--1557},
  year={2025, Rotterdam, Netherlands}
}

@article{sv3,
  title={Advancing Zero-Shot Open-Set Speech Deepfake Source Tracing},
  author={Chhibber, Manasi and Mishra, Jagabandhu and Kinnunen, Tomi H},
  journal={arXiv preprint arXiv:2509.24674},
  year={2025, \{Last Accessed: $27^{th} February, 2026$\}}
}

@article{lecun2015deep,
  title={Deep learning},
  author={LeCun, Yann and Bengio, Yoshua and Hinton, Geoffrey},
  journal={Nature},
  volume={521},
  number={7553},
  pages={436--444},
  year={2015},
  publisher={Nature Publishing Group UK London}
}

@inproceedings{siami2019performance,
  title={The performance of LSTM and BiLSTM in forecasting time series},
  author={Siami-Namini, Sima and Tavakoli, Neda and Namin, Akbar Siami},
  booktitle={IEEE International Conference on Big Data (Big Data)},
  pages={3285--3292},
  year={2019, Los Angeles, CA, USA},
  organization={}
}

@inproceedings{cho2014learning,
  title={Learning phrase representations using RNN encoder--decoder for statistical machine translation},
  author={Cho, Kyunghyun and Van Merri{\"e}nboer, Bart and Gul{\c{c}}ehre, {\c{C}}a{\u{g}}lar and Bahdanau, Dzmitry and Bougares, Fethi and Schwenk, Holger and Bengio, Yoshua},
  booktitle={Conference on Empirical Methods in Natural Language Processing (EMNLP)},
  pages={1724--1734},
  year={2014, Doha, Qatar}
}

@inproceedings{pesq,
  title={Perceptual evaluation of speech quality ({PESQ})-a new method for speech quality assessment of telephone networks and codecs},
  author={Rix, Antony W and Beerends, John G and Hollier, Michael P and Hekstra, Andries P},
  booktitle={IEEE International Conference on Acoustics, Speech, and Signal Processing (ICASSP)},
  volume={2},
  pages={749--752},
  year={2001, Salt Lake City, Utah, USA},
  organization={}
}

@article{stoi,
  title={An algorithm for intelligibility prediction of time--frequency weighted noisy speech},
  author={Taal, Cees H and Hendriks, Richard C and Heusdens, Richard and Jensen, Jesper},
  journal={IEEE Transactions on Audio, Speech, and Language Processing},
  volume={19},
  number={7},
  pages={2125--2136},
  year={2011},
  publisher={}
}

@incollection{pcc,
  title={{Pearson Correlation Coefficient}},
  author={Benesty, Jacob and Chen, Jingdong and Huang, Yiteng and Cohen, Israel},
  booktitle={Noise Reduction in Speech Processing},
  pages={1--4},
  year={2009},
  publisher={Springer}
}

@article{msd,
  title={{RASTA} processing of speech},
  author={Hermansky, Hynek and Morgan, Nelson},
  journal={IEEE Transactions on Speech and Audio Processing},
  volume={2},
  number={4},
  pages={578--589},
  year={2002},
  publisher={}
}

@inproceedings{MCD,
  title={Mel Cepstral Distance measure for objective speech quality assessment},
  author={Kubichek, Robert},
  booktitle={IEEE Pacific Rim Conference on Communications Computers and Signal Processing (PACRIM)},
  volume={1},
  pages={125--128},
  year={1993, Victoria, BC, Canada},
  organization={}
}

@article{mos,
  title={{Mean Opinion Score (MOS)} revisited: Methods and applications, limitations and alternatives},
  author={Streijl, Robert C and Winkler, Stefan and Hands, David S},
  journal={Multimedia Systems},
  volume={22},
  number={2},
  pages={213--227},
  year={2016},
  publisher={Springer}
}

\bibliographystyle{abbrv}
\appendix
\section{Performance Metrices}
\label{appendix_b}
\subsection{Objective Evaluation Metrics}

\begin{enumerate}
    \item \textbf{Perceptual Evaluation of Speech Quality (PESQ) \cite{pesq}:}
    
    This metric estimates how close the processed speech is to the original. Higher PESQ scores indicate better quality. It can be calculated as:
    
    \begin{equation}
    \mathrm{PESQ} = 4.5 - 0.1d_{\mathrm{sym}} - 0.0309d_{\mathrm{asym}}
    \end{equation}
    
    where $d_{\mathrm{sym}}$ captures the effect of symmetric distortions (differences between clean and degraded signals), and $d_{\mathrm{asym}}$ reflects asymmetric distortions (artifacts that affect one side more than the other).

    \item \textbf{Short-Time Objective Intelligibility (STOI) \cite{stoi}:}
    
    STOI measures how understandable speech is under noisy or degraded conditions. The score ranges from 0 to 1, where values closer to 1 indicate higher intelligibility. It is defined as:
    
    \begin{equation}
    \mathrm{STOI}
    =
    \frac{1}{N}
    \sum_{t=1}^{N}
    \frac{
    \sum_{f=1}^{F}
    \left(x_f(t)-\bar{x}_f\right)
    \left(y_f(t)-\bar{y}_f\right)
    }{
    \sqrt{
    \sum_{f=1}^{F}
    \left(x_f(t)-\bar{x}_f\right)^2
    }
    \sqrt{
    \sum_{f=1}^{F}
    \left(y_f(t)-\bar{y}_f\right)^2
    }
    }
    \end{equation}

    \item \textbf{Pearson Correlation Coefficient (PCC) \cite{pcc}:}
    
    PCC evaluates the linear correlation between two signals $x$ and $y$. The score ranges from $+1$ (perfect positive correlation) to $-1$ (perfect negative correlation). It is calculated as:
    
    \begin{equation}
    \mathrm{PCC}
    =
    \frac{
    \sum_{i=1}^{N}
    (x_i-\bar{x})(y_i-\bar{y})
    }{
    \sqrt{
    \sum_{i=1}^{N}
    (x_i-\bar{x})^2
    }
    \sqrt{
    \sum_{i=1}^{N}
    (y_i-\bar{y})^2
    }
    }
    \end{equation}

    \item \textbf{Modulation Spectra Distance (MSD) \cite{msd}:}
    
    MSD quantifies the similarity between the modulation spectra of two signals. A lower MSD indicates the synthesized signal is closer to the reference. It is given by:
    
    \begin{equation}
    \mathrm{MSD}
    =
    \sqrt{
    \frac{1}{N}
    \sum_{i=1}^{N}
    \left(
    s(y)_i^{t}
    -
    s(y)_i^{\hat{t}}
    \right)^2
    }
    \end{equation}
    
    where $s(y)_i^{t}$ and $s(y)_i^{\hat{t}}$ represent the $i^{\text{th}}$ modulation spectral component of the reference and synthesized signals, respectively.

    \item \textbf{Mel-Cepstral Distortion (MCD) \cite{MCD}:}
    
    MCD measures the difference between Mel-cepstral coefficients of two signals. A smaller value indicates the synthesized speech is more similar to the ground truth. It is calculated as:
    
    \begin{equation}
    \mathrm{MCD}(k)
    =
    \sqrt{
    \sum_{i=1}^{16}
    \left[
    C_x(i,k)-C_y(i,k)
    \right]^2
    }
    \end{equation}

\end{enumerate}

\subsection{Subjective Evaluation Metric}

\begin{enumerate}
    \item \textbf{Mean Opinion Score (MOS) \cite{mos}:}
    
    MOS is based on human perception, where listeners rate the quality of speech on a scale of 1 to 5. The final score is the average of all listener ratings:
    
    \begin{equation}
    \mathrm{MOS}
    =
    \frac{\sum_{n=1}^{N} R_n}{N}
    \end{equation}
    
    where $R_n$ is the rating provided by the $n^{\text{th}}$ listener and $N$ is the total number of listeners.
    
\end{enumerate}
\section{Language distribution}
\label{appendix_a}
\begin{figure*}[h]
    \centering
    \includegraphics[width=\textwidth]{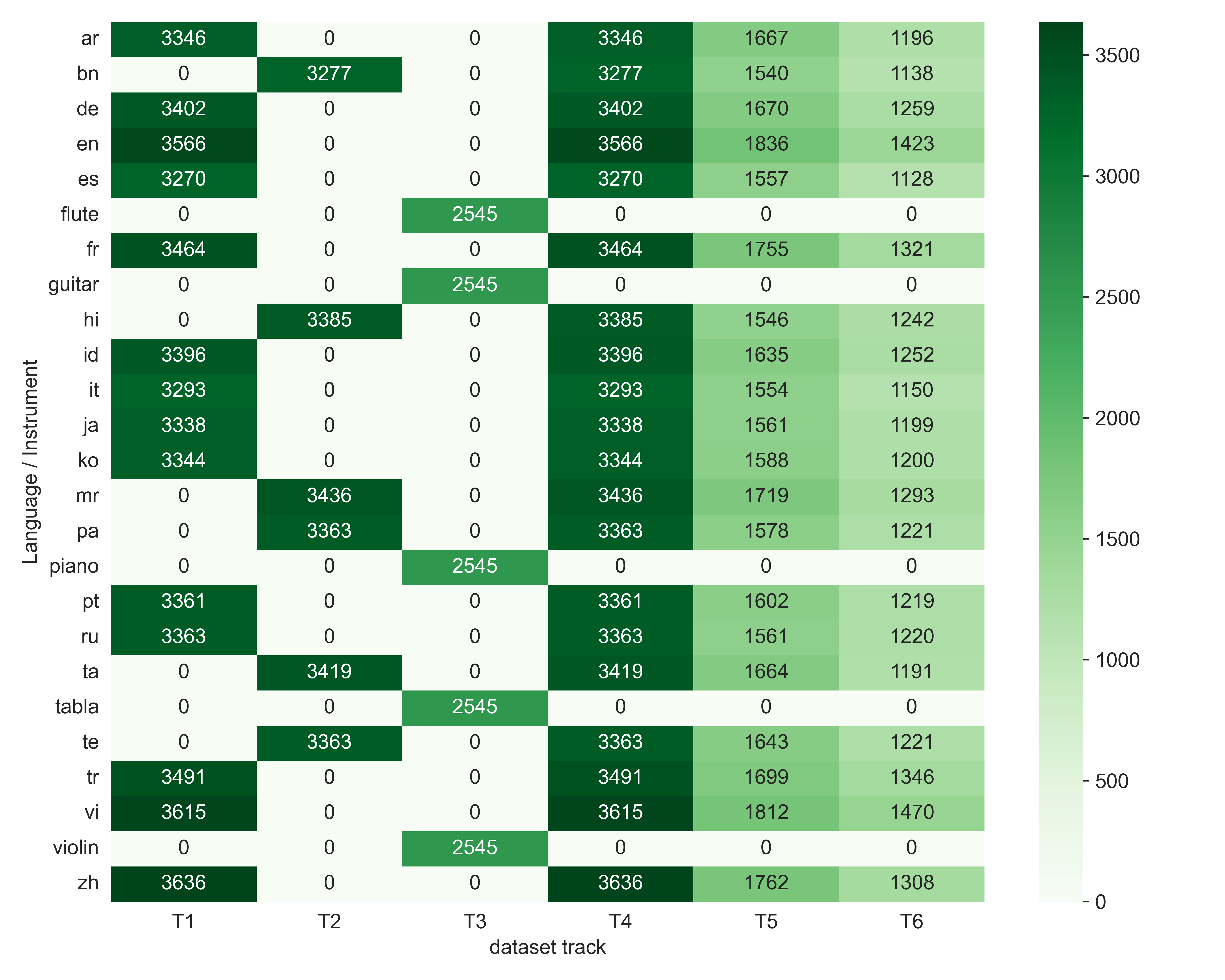}
    \caption{Distribution on each language, and instrument in each track of dataset.}
    \label{fig:language_files_per_track}
\end{figure*}
\end{document}